# Design and Implementation of a Procedural Content Generation Web Application for Vertex Shaders

Juan C. Quiroz
Computing and Information Systems
Sunway University
Bandar Sunway, Malaysia
juanq@sunway.edu.my

Sergiu M. Dascalu
Computer Science and Engineering
University of Nevada, Reno
Reno, NV, USA
dascalus@cse.unr.edu

**Abstract** *We present a web application for the procedural generation of transformations of 3D models. We generate the transformations by algorithmically generating the vertex shaders of the 3D models. The vertex shaders are created with an interactive genetic algorithm, which displays to the user the visual effect caused by each vertex shader, allows the user to select the visual effect the user likes best, and produces a new generation of vertex shaders using the user feedback as the fitness measure of the genetic algorithm. We use genetic programming to represent each vertex shader as a computer program. This paper presents details of requirements specification, software architecture, high and low-level design, and prototype user interface. We discuss the project's current status and development challenges.*

## 1 Introduction

Procedural content generation (PCG) is the algorithmic creation of game content. PCG provides the potential to reduce the cost and time to create content, while also augmenting the creativity of designers, artists, and programmers [1]. We present a PCG web application that enables users to create transformations of 3D models. Rather than creating 3D models from scratch or from a set of polygonal primitives, we start with a well-formed 3D model and explore variations of the 3D model. The transformations are implemented with vertex shaders.

A vertex shader allows mathematical operations to be performed on the individual vertices that make up a 3D model [2]. The vertex shader performs operations on each vertex, and thus provides great flexibility to modify the position, color, texture, and lighting of individual vertices. The challenge is that making a vertex shader requires programming and computer graphics experience. In addition, even if a programmer writes a vertex shader that produces interesting results, creating a variation of that vertex shader is not straightforward.

We present an interactive genetic algorithm (IGA) to allow users to explore transformations of 3D models. In an IGA, a user guides a search process by visually evaluating solutions and providing feedback based on personal preferences [3]. In our web application, we present the user with transformations of a 3D model, and the user selects the transformations he/she likes the best. The user feedback is used by the IGA to generate new transformations, which are then presented to the user for evaluation, with the process repeating until the user is satisfied. The IGA generates the vertex shaders using genetic programming (GP) [4]. In GP, computer programs are typically represented as tree structures [4], [5].

This paper makes two contributions. First, we present a web PCG application for exploring transformations of 3D models. This system allows users to create transformations without having to install plug-ins or any additional software on their computers. Our second contribution is, we use GP to evolve vertex shaders which perform operations on the position of the vertices that make up the 3D models. We describe the specification and design of our application.

The remainder of this paper is structured as follows. Section 2 describes background on genetic algorithms and related work. Section 3 lists the functional and nonfunctional software requirements for the system. Section 4 presents the use case diagram and the use cases of the system. Section 5 describes the system's architecture. Section 6 reports on the system's current status and future work.

## 2 Background

Our web application uses an existing 3D model as the seed to explore variations of the 3D model with an interactive genetic algorithm (IGA). The user thus explores and evaluates vertex shaders by seeing the rendered result of each vertex shader applied to the 3D model and guiding the

IGA via subjective feedback. Our implementation relies on genetic algorithms (GAs), interactive genetic algorithms (IGAs), and genetic programming (GP) to create the vertex shaders. We describe each of these next.

## 2.1 Genetic Algorithms and Genetic Programming

A GA is a search algorithm based on the principles of genetics and natural selection [5]. The GA maintains a population of individuals, where each individual is a potential solution to the problem being solved. In our case, each individual consists of a vertex shader program. During initialization of the GA, the individuals are created randomly. To generate a new population, parents are selected for crossover, with parent selection favoring the fittest individuals. The resulting offspring are mutated. This process repeats until a terminating condition is met.

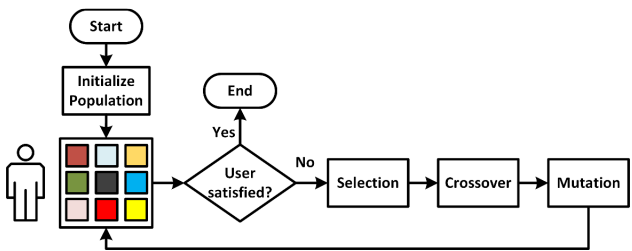

Figure 1: Flowchart of interactive genetic algorithm.

In a GA, fitness evaluation assigns a fitness value to each individual in the population. This is what drives the search process towards higher fitness solutions. When dealing with subjective criteria, such as aesthetics and user preferences, it can be difficult to create an algorithm to determine the fitness of an individual [3]. An IGA solves this by replacing the fitness evaluation with user evaluation, as illustrated in Figure 1. Thus, the IGA presents the user with individuals from the population, the user evaluates the solutions—scores, ranking, selecting the best—and the IGA proceeds with selection, crossover, and mutation.

In the canonical GA, individuals in the population are encoded using a binary string representation. In genetic programming, computer programs are encoded and evolved by the GA [4]. The computer programs are typically represented as tree structures. In our web application we use GP to create mathematical equations to change the positions of vertices in the vertex shader.

## 2.2 Related Work

Evolutionary computation has been used in prior work for evolving vertex shaders [6]–[9]. In [6], linear GP is used to evolve vertex and pixel shaders. However, their work is limited to applying texture, color, and lighting values of pixels and vertices, and they do not change the positions of vertices. In [7], GP is used to create vertex shaders to apply coloring schemes to scenes being rendered. In [10], Meyer-Spradow et al. use GP to evolve fragment shaders that apply materials to 3D objects. In [11], GP is used to evolve high-level Renderman shaders. In [12], an unsupervised approach is used to generate 2D textures, with GP evolving texture images that are similar to one or more target images.

In our prior work, we used an IGA to evolve equations in vertex shaders to create transformations of 3D models [8], [9]. Our prior IGA implementation used Python-Ogre, the python binding for the OGRE graphics rendering engine. The limitations of our prior implementation were that (1) it was a desktop application requiring users to install a large set of libraries to run the program, (2) it used bit-encoded, full binary tree representation for the individuals in the IGA. In particular, the bit-encoded, full binary tree representation was a limitation imposed given the requirements of the prior project. However, it was clear that an unrestricted GP implementation was needed to explore a wide range of equations for vertex shaders. Our new implementation uses GP for the encoding of the vertex shaders, allowing the IGA to explore a bigger space of transformations.

IGAs have also been used for creating 3D models, without using GP or vertex shaders. In [13], Huang et al. used an IGA to create 3D avatars by evolving the values of parameters that defined the properties of the avatar. In [14], 3D model shapes were created with the implicit surface method, which blended primitive shapes into a complex shape. In [15] an IGA evolved the parameters used by a procedural tree generation algorithm to create 3D trees.

The design and implementation described in this paper provides a web application to create and explore transformations of 3D models right on the user's web browser. There is no need to install any additional software or plugins on the client computer. Our work is also the first to use GP to apply operations that change the positions of the vertex data. In contrast, prior GP has been used to perform operations on vertex data other than position, such as colors, textures, and materials.

## 3 Requirements Specification

### 3.1 Functional Requirements

The functional requirements describe the most important behavior of the software, including user interactions and rendering processes.

1. The system shall render a 3D model.
2. The system shall divide the screen into a 3x3 grid, with a viewport for each cell in the grid.
3. The system shall use a camera for each viewport.

4. The system shall display a 3D model within each viewport.
5. The system shall allow the user to move all of the cameras simultaneously with a single set of keyboard controls and the mouse.
6. The system shall allow the user to select with the mouse the 3D model the user likes the best.
7. The system shall allow the user to step the IGA a number of generations.
8. The system shall allow the user to save a selected transformation of a 3D model.
9. The system shall allow the user to start the IGA.
10. The system shall allow the user to load a 3D model from the file system.
11. The system shall allow the user to browse transformations stored in a public database.
12. The system shall allow the user to use a transformation loaded from a public database to seed the IGA.

## 3.2 Non-Functional Requirements

The non-functional requirements outline the most important constraints on the system.

1. The system shall render in real-time.
2. The system shall be implemented with HTML, JavaScript, and Three.js
3. The system shall run on a web browser with HTML5 support and hardware acceleration.
4. The server request handling shall be implemented with Python.
5. The IGA shall be implemented with the Distributed Evolutionary Algorithms in Python library.
6. The system shall be implemented using a representational state transfer (REST) architecture.
7. The system shall use GP to generate the vertex shader equations.
8. The system shall use 3D models in JSON format.
9. The system shall use time as a variable in the vertex shader equations.

# 4 Use Case Modeling

To gain further insight into the functionality of the system, we have divided the behavior of the web application into use cases. The use diagram in Figure 2 outlines the controls that allow the user to interact with the system, and the functionality on the server-side for running the IGA. In order to further clarify the functionality, detailed descriptions of each use case are presented next.

**UC01** *Start Evolution:* The user selects the parameters of the IGA. The user then pushes the start button to display the first set of transformations.

**UC02** *Select Model:* The user selects the transformation the user likes best. The user can select multiple transformations.

**UC03** *Step Generation:* The user submits the selected transformation by clicking the Next button or by using a keyboard shortcut. This sends a request to the server to assign fitness values to the individuals in the population and create a new population.

**UC04** *Camera Control:* The user can move the camera to zoom in to the 3D model or to view the 3D model from a different angle. The camera controls will move all of the viewport cameras at the same time. The keyboard and the mouse can be used to control the camera.

**UC05** *Load Model:* The user can upload a 3D model in JSON format using a file browser dialog. After the upload completes, the 3D model will be displayed within each viewport. The user can also browse 3D models uploaded by other users, select one of the models, and load the model to the scene.

**UC06** *Save Transformation:* The user can save a transformation by first making a selection and then clicking the save button.

**UC07** *Load Transformation:* The user can load a transformation from a file or from a database of transformations created collaboratively by users over time. Loading the transformation will also inject the corresponding vertex shader equation to the IGA population.

**UC08** *Render Scene:* The GPU on the client device will render the scene, applying a vertex shader to each of the 3D models.

**UC09** *Initialize Population:* The server initializes the population of the IGA in response to receiving a request to start the evolutionary process.

**UC10** *Fitness Evaluation:* The selections from the user are received and used to calculate the fitness of the individuals in the IGA population. The fitness of each individual is calculated by determining the similarity between the individual and the user selections.

**UC11** *New Population*: The IGA performs selection, crossover, and mutation to generate a new population of individuals.

**UC12** *Select Subset:* Out of the large population size of the IGA, the server selects a subset of vertex shaders to return to the client for rendering on the user's screen.

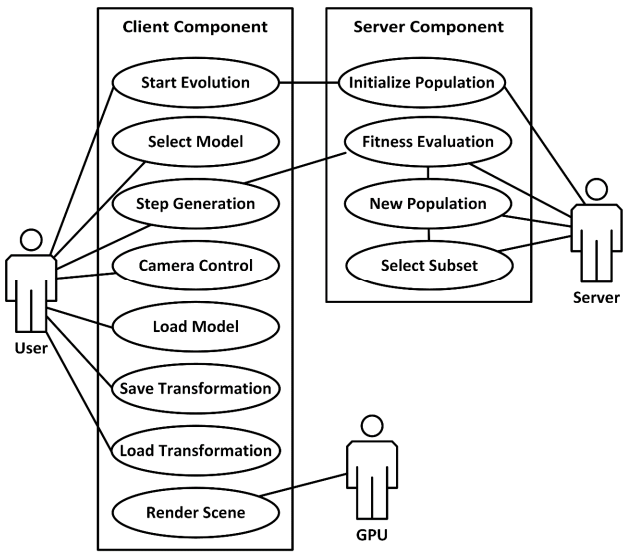

Figure 2: Use case diagram for web application.

# 5 Architectural and Detailed Design

## 5.1 Architectural Design

Figure 3 illustrates the architecture of the web application. The application consists of a front-end and a back-end. The front-end also includes all of the graphics rendering on the web browser. WebGL is a JavaScript API for rendering 2D and 3D graphics within compatible browsers. Most importantly, no plug-ins need to be installed for the rendering to work. We use Three.js, a 3D JavaScript library, which simplifies the process of writing the WebGL code.

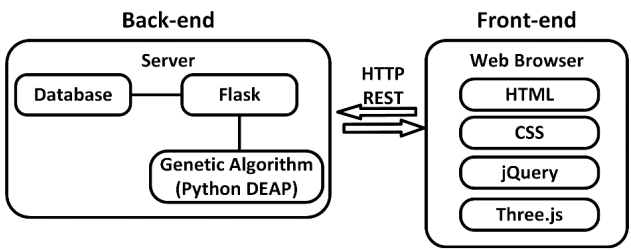

Figure 3: RESTful architecture for web application.

The back-end includes all the server processing. We used the REST architecture for the communication protocol between client and server. We implemented the RESTful web services with Flask, a web microframework for Python. Our RESTful API provides interaction to the IGA for initializing the population, receiving the user selections, generating a new population, saving transformations, and loading transformations and models. The IGA was implemented using the Distributed Evolutionary Algorithms in Python (DEAP) library. The database stores the vertex shaders saved by the user and uploaded models. In our current implementation, the database has not been completed. We leave the database, which would provide the ability for users to collaboratively build a library of vertex shaders, for future work.

## 5.2 System Activity Chart

Figure 4 presents an activity chart of the web application, showing the interaction between the user and the IGA. The user starts the evolution, which sends a GET request to the server to retrieve vertex shaders to apply to the models currently rendered on the user's web browser. The IGA maintains a large population size, and from this large population, a subset is selected to be evaluated by the user [9]. The user then has the option of selecting a model, and submitting this as fitness feedback to the IGA, with fitness evaluation done as in [9]. The user can repeat this process for as many generations as he/she wants.

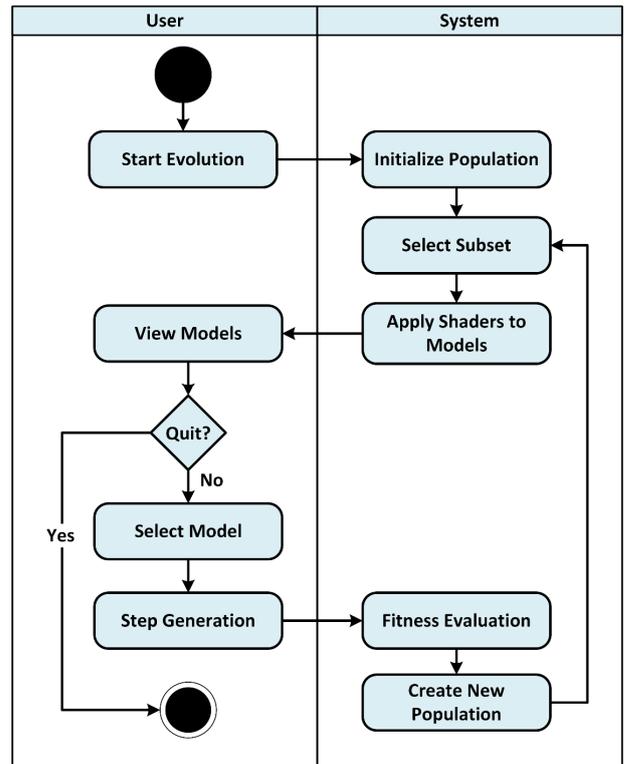

Figure 4: Activity chart for user interaction IGA.

## 5.3 Low Level Design

Figure 5 illustrates an example tree structure generated using GP. The tree structure represents the expression: $x / (x + z)$, where x and z are the x and z values of the current vertex. In our implementation, the non-leaf nodes include the operators of addition, subtraction, multiplication, division, negation, and the trigonometric functions of sine and cosine. The terminals include constants uniformly distributed over the half-open interval [-1, 1), a time variable, and the x, y, or z coordinate of the current vertex. The program, such as the one in Figure 5, is evaluated and it results on a scalar value. This scalar value is added to the x, y, and z coordinates of the current vertex. For example, the equation from Figure 5 would be used in the vertex shader as shown in equation (1), where "*position.xyz*" represents a vector containing the x, y, and z coordinates of the current vertex.

$$position.xyz = position.xyz + x/(x+z) \qquad (1)$$

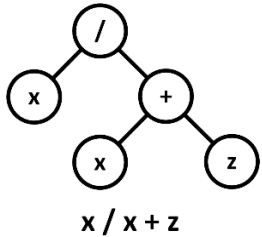

Figure 5: Sample tree structure and its evaluation.

## 6 Current Status and Future Work

Figure 6 shows the interface of the web application when the web page is initially loaded. The system loads a default 3D model, consisting of a human character. A button is provided next to each model to select a model. The scene is divided into viewports, forming a 3x3 grid. Figure 7 shows a screenshot after 8 generations. Some of the transformations result in aesthetically pleasing effects, while other transformations are quite destructive. For instance, the transformations in the middle column in Figure 7 are not practical for video game use. In some cases, the equations do not generate noticeable effects, such as the model in row two – column one.

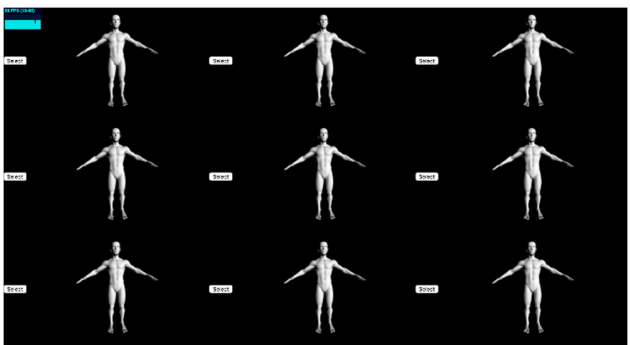

Figure 6: Main interface of the web application.

The use of the time variable makes a strong difference in the resulting transformations. If the equation in the vertex shader includes the time variable, the transformation is dynamic. If the equation does not include the time variable, the transformation is static. We leave the analysis of static versus dynamic transformations for future work.

Figure 8 illustrates various transformations generated with our web application. Figures 8(a)-(e) illustrate aesthetically pleasing transformations, where the shape of the 3D model can still be appreciated. The use of trigonometric functions tends to result in curvy models, giving the models a more finished appearance. The transformations in Figure 8(f)-(h) are quite destructive as the shape of the original model is entirely lost, making the transformations look like noise.

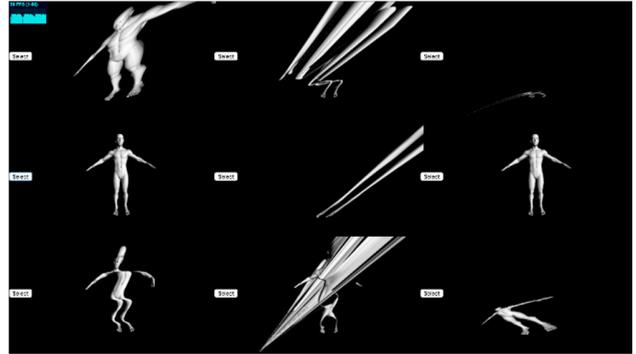

Figure 7: Main interface after evolution has started.

A limitation of our current implementation is the user interaction. IGAs tend to suffer from user fatigue, user boredom, and user inconsistency in the input provided to the GA [3]. Our IGA uses a feedback method as discussed in [9], where the user simply selects the models he/she likes best from a small subset sampled from the large population size of the IGA. Further work is needed to make the user interaction engaging and intuitive to users from generation to generation. A second limitation is the type of equations that can be generated with our GP implementation based on our selection of operators and terminals. The format of equation (1) also introduces a strong limitation on how the GP equations can modify the vertex data, since the x, y, and z coordinates of each vertex are modified by the same expression.

Finally, our current implementation uses least square sums to determine the similarity between the user selected vertex shader and the individuals in the IGA. That is, least square sums is used to determine GP tree similarity. However, early testing has shown that this similarity metric does not make the IGA exploration intuitive. We leave the analysis of various similarity metrics for future work.

## 7 Conclusions

We presented a PCG system for evolving transformations of 3D models. The transformations are encoded using GP, with an IGA allowing the user to quickly explore transformations based on his/her preference. Our PCG system uses a web application implementation, allowing users to create transformations on their web browser, without having to install libraries or plug-ins.

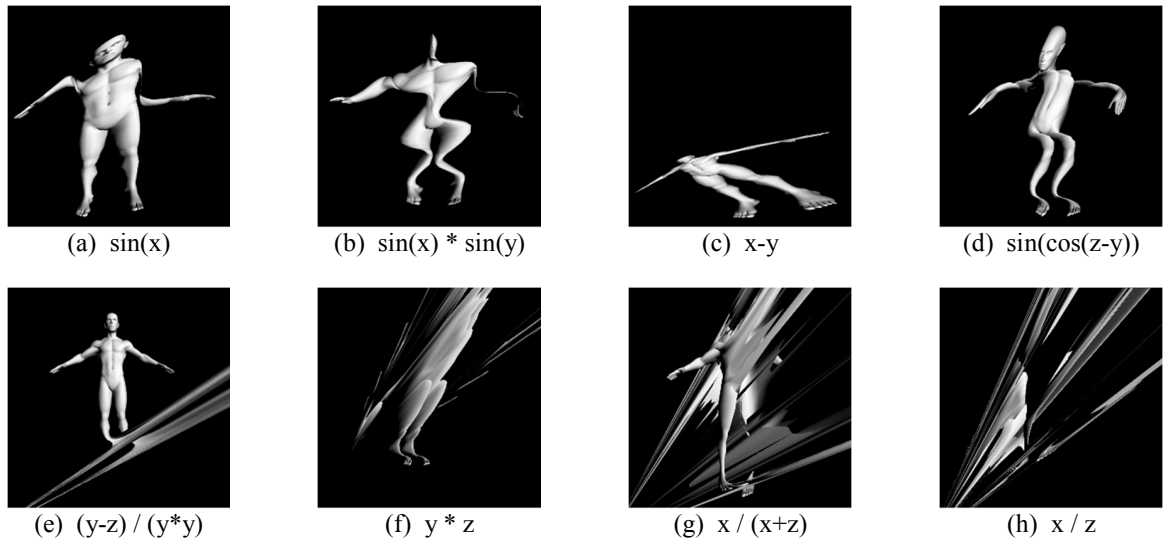

(a) sin(x)    (b) sin(x) * sin(y)    (c) x-y    (d) sin(cos(z-y))

(e) (y-z) / (y*y)    (f) y * z    (g) x / (x+z)    (h) x / z

Figure 8: Example transformations generated by the corresponding equation in the vertex shader.